\documentclass[12pt,preprint]{aastex}

\newcommand{\etal}{{\it et al. \/}}

\newcommand{\msin}{M{\rm sin}\;i}
\newcommand{\vsin}{v{\rm sin}\;i}

\begin{document}
\title{What Fraction of Sun-like Stars have Planets?}

\medskip

\author{Charles H. Lineweaver \& Daniel Grether \\
University of New South Wales \\
charley@bat.phys.unsw.edu.au}


\begin{abstract}
The radial velocities of $\sim 1800$ nearby Sun-like
stars are currently being monitored by eight high-sensitivity Doppler exoplanet surveys. 
Approximately 90 of these stars have been found to host exoplanets massive enough to be detectable. 
Thus at least $\sim 5\%$ of target stars possess planets.
If we limit our analysis to target stars that have been monitored the longest ($\sim 15$ years), $\sim 11 \%$ possess planets. 
If we limit our analysis to stars monitored the longest and whose low surface activity allow
the most precise velocity measurements,   
$\sim 25 \%$ possess planets.
By identifying trends of the exoplanet mass and period distributions in a sub-sample of exoplanets less-biased by 
selection effects, and linearly extrapolating these trends into regions 
of parameter space that have not yet been completely sampled, we find at least 
$\sim 9\%$ 
of Sun-like stars have planets in the mass and orbital period ranges $\msin > 0.3 M_{Jupiter}$ and $P < 13\; \mbox{\rm years}$, 
and at least $\sim 22\%$ 
have planets in the larger range $\msin > 0.1 M_{Jupiter}$ and $ P < 60\; \mbox{\rm years}$.
Even this larger area of the log mass - log period 
plane is less than $20\%$ of the area occupied by our planetary system,
suggesting that this estimate is still a lower limit to the true fraction of Sun-like stars with planets, which may be 
as large as $\sim 100\%$.
\end{abstract}
\section{Introduction}

With increasingly sensitive instruments exoplanet hunters have detected more than 100 exoplanets.  
The focus of these pioneering efforts has been to find and describe new exoplanets.
As more exoplanets have been found, the question: `What fraction of stars have planets?' has been looked at periodically.
Estimates of the fraction of stars with planets can be simply calculated from the raw numbers of exoplanet hosts divided by the
number of monitored stars. For example, Marcy \& Butler (2000) 
report ``5\% harbor companions of $0.5$ to $8 M_{Jupiter}$ within $3$ AU". 
Estimates can also be based on high precision Doppler targets in a single survey. 
Fischer \etal (2003) 
report a fraction of $15\%$ for high precision Doppler targets in the original Lick sample.
Estimates can also be based on a semi-empirical analysis of exoplanet data (Tabachnik and Tremaine 2002).

In this work we use a semi-empirical method, staying as close to the exoplanet data as possible.
We use the growth and current levels of exoplanet detection to verify that sensitive and long duration 
surveys have been finding more planets in a predictable way.
This is no surprise. However, quantitatively following the consistent increase of the lower limit to the 
fraction of stars with detected planets is an important new way to substantiate both current and future 
estimates of this fraction. 
Because of the growing importance of the question `What fraction of stars have planets?' and our 
increasing ability to answer this question, such a closer scrutiny of (i) the assumptions used to arrive 
at the answer, (ii) the parameter space in which they are valid and (iii) the associated error bars, is timely. 

If the Sun were among the
target stars of the Doppler surveys, it would be another few years
before the presence of Jupiter (the most easily detected feature of our planetary system) could be confidently detected. 
However, we are beginning to be able to answer the question: How typical is Jupiter? 
In Lineweaver \& Grether (2002) we found that extrapolations of 
trends found in a less-biased sub-sample of exoplanet data indicate that Jupiter is a typical massive planet in the sense 
that it lies in the most densely occupied region of the log mass - log period plane. In Lineweaver \etal (2003) we 
updated our analysis to include exoplanets detected between January and August 2002 and found more support for this
idea. 
However, in this previous work we made no effort to determine the absolute frequency of Jupiters or to answer the question: 
What fraction of stars have planets? That is the main focus of this paper.

\begin{figure}[!h]
\epsscale{0.9}
\plotone{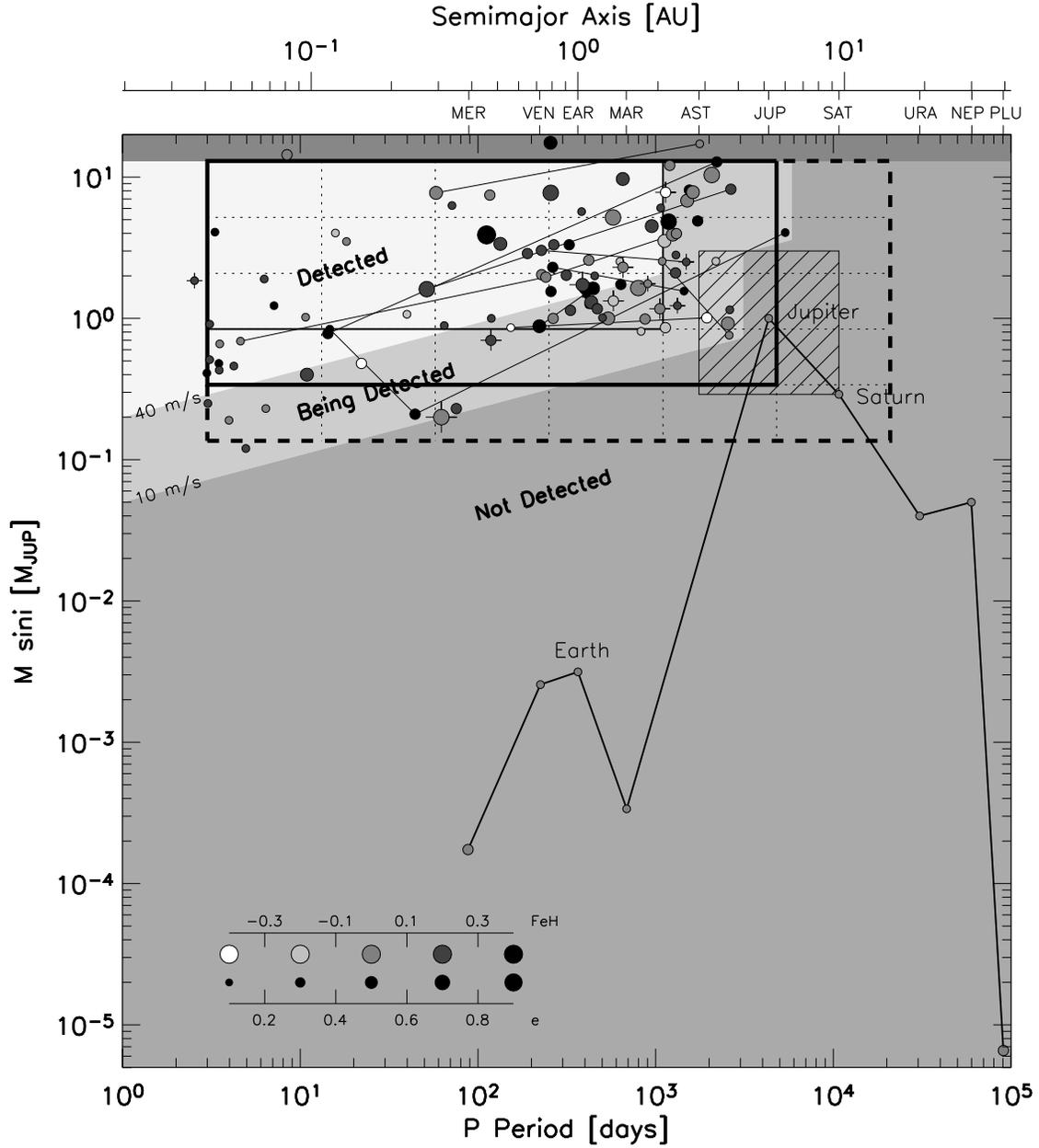}
\caption{Our Solar System compared to the 106 exoplanets detected by eight high-sensitivity 
Doppler surveys.
The three regions labeled ``Detected'', ``Being Detected'' and ``Not Detected'' indicate the selection 
effects in the log mass - log period plane due to a limited time of observation and limited radial velocity sensitivity
(Section \ref{sec:mpdist}). 
The rectangular region circumscribed by the thin solid line, fits almost entirely into the ``Detected'' region, and
contains what we call the less-biased sub-sample. This sub-sample is the basis of the trends in mass and period identified in 
Fig. \ref{fig:EP_Hist_log_lin}.
Metallicity of the host stars are indicated by the shade of gray of the point while eccentricity of the exoplanet 
orbits are indicated by point size (see key in lower left). Thin lines connect planets orbiting the same host star. 
The cross-hatched square is our Jupiter-like region defined by $M_{Saturn} \leq \msin \leq 3 M_{Jupiter}$ 
and $P_{asteroids} \leq P \leq P_{Saturn}$. 
}
\label{fig:EP_m_p_big}
\end{figure}

In Section \ref{sec:exoplanetdata} we present the exoplanet data set (Fig. \ref{fig:EP_m_p_big}). 
We analyze the target lists and detections and show how
the fraction of stars with detected planets has increased over time 
(Figs. \ref{fig:EP_Hist_pyramid} \& \ref{fig:EP_Hist_timeline}).
In Section~\ref{sec:highDopp} we quantify how the fraction of stars with planets 
depends on the precision with which the radial velocity of individual stars can be measured (Fig. \ref{fig:highDoppler}).
In Section~\ref{sec:slopes} we quantify trends in exoplanet mass and period by linear and power-law fits
to histograms of a less-biased sub-sample of exoplanets (Fig.~\ref{fig:EP_Hist_log_lin}).
Based on these trends, we extrapolate into larger regions of parameter space
and give estimates for the fraction of stars with planets in these regions.
In Section \ref{sec:summary} we summarize, discuss and compare our results with previous work.
Our approach is most similar and complimentary to the work of Fischer \etal (2003) and 
Tabachnik \& Tremaine (2002).

\clearpage

\clearpage
\section{Exoplanet Data}
\label{sec:exoplanetdata}


\subsection{Mass-period Distribution}
\label{sec:mpdist}
Figure 1 displays the masses and periods of the 106 exoplanets detected as of June 2003 by the eight high precision
Doppler surveys analyzed in this paper.
The region in the upper left labeled ``Detected'' is our estimate of the region in which 
the Doppler technique has detected virtually all the exoplanets with periods less than
three years orbiting target stars that have been monitored for at least three years.
This region is bounded by three days on the left, three years on the right, $13 M_{Jupiter}$ on the top and at the bottom by
a radial velocity of $40$ m/s induced in a solar-mass host star.
The largest observed exoplanet period and the smallest observed radial velocity induced by a detected exoplanet are used to
define the boundary between the ``Being Detected'' and ``Not Detected'' regions.
The discontinuity of the ``Being Detected'' region near Jupiter is due to the increased sensitivity of the 
original Lick survey at the end of 1994.
No more exoplanets should be detected in the ``Detected'' region unless new stars are 
added to the target lists. Thus, all the exoplanets marked with a `+' (detected since August 2002) should fall in,
or near, the ``Being Detected'' region.
Of the 11 new detections since August 2002, 8 fall in the ``Being Detected'' region, 2 fall just inside the ``Detected'' region 
while one has $P < 3$ days -- it was not being monitored with sufficient phase coverage for detection until recently.

We define a less-biased sample of planet hosts as the 49 hosts to the planets within the 
rectangular region circumscribed by a thin solid line ($3$ days $< P < 3$ years and $0.84 < \frac{\msin}{M_{Jupiter}} < 13$)
in Fig. \ref{fig:EP_m_p_big}.
This rectangle is predominantly in the ``Detected'' region.  We will use this less-biased sample as 
the basis for our extrapolations (Sec. \ref{sec:slopes}).  As a completeness correction for the lower 
right corner of this rectangle being in the `Being Detected' region,  we add 4 planets as described 
in \cite{LGH03} giving us a corrected less-biased sample of $53 (=49 + 4)$ planets.

\subsection{Increasing Fraction as a function of monitoring duration}
\label{sec:time}

\begin{figure}
\vspace{-0.5cm}
\epsscale{0.8} 
\plotone{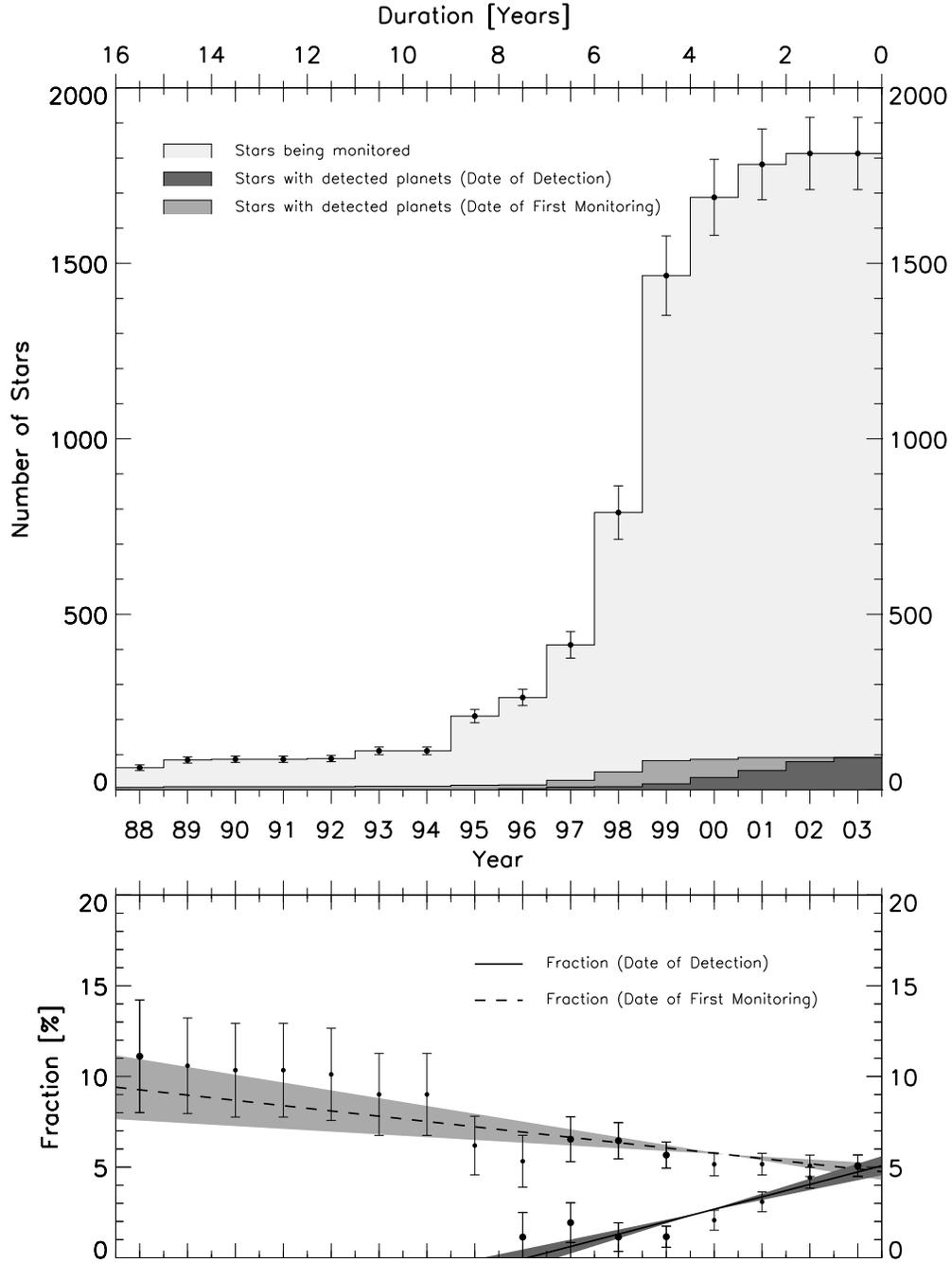}
\caption{ \scriptsize
The fraction of stars with planets.
TOP: Histograms of the increasing number of target stars and the increasing number of them found 
to be hosting at least one exoplanet. 
The host stars are binned in two ways. In the darkest histogram they are binned at 
the date of detection of the first exoplanet found orbiting the star. In the middle-gray histogram they are binned at 
the date of first monitoring of the star. Notice the small number of target stars until the discovery of the first 
planet (Mayor \& Queloz 1995). 
BOTTOM: By taking the ratio, in the top panel, of the darker histograms to the 
lightest one, we obtain the fraction of target stars hosting at least one planet. 
The two binning conventions lead to different results. Using the date-of-detection binning (darkest histogram) 
yields the intuitive result that the fraction of host stars starts at zero in 1995 and climbs monotonically until its 
current value of $5\pm 1\% \; (\approx 92/1812)$. The date-of-first-monitoring binning starts on the far left at 
$11 \pm 3\%$ from the bin of target stars that has been monitored the longest. 
The fraction decreases as we average in more stars 
that have been monitored for shorter durations. The dashed line is the result of a linear fit to a sparse, and therefore
more independent, sample of the non-independent points (data points used have a larger point size). 
This line yields $9 \pm 2 \%$ on the far left in good
agreement with the single data point in the left-most bin. 
The last bin on the right is the same in both binning methods and includes all Sun-like target stars. Thus, $\approx 5\%$  
is an average fraction from target stars that have been monitored for times varying between 0 and 16 years, 
while $\approx 11\%$ is the fraction from target stars that have been monitored the longest ($\approx 15$ yrs). 
The non-trivial task of estimating the total number of stars being monitored is described in Appendix A and 
summarized in Table 4.  
}
\label{fig:EP_Hist_pyramid}
\end{figure}

The exoplanets plotted in Fig. 1 are the combined detections of
eight Doppler surveys  currently monitoring the radial 
velocities of $\approx 1812$ nearby FGK stars (Tables 3 $\&$ 4).
The top panel of Figure \ref{fig:EP_Hist_pyramid} shows how the number of these target stars has increased over 
the past 16 years. 94 stars have been found to host 106 exoplanets.
Of these 94, 92  fall within our selection criteria of Sun-like stars (= FGK class IV or V) 
with planets ($\msin < 13 M_{Jupiter}$). 
%
Six known exoplanet hosts were not included in this analysis because they were found in the context 
of surveys whose search strategies and sensitivities cannot easily be combined with results from the 
8 sensitive Doppler surveys analyzed here.  

%


In Fig. \ref{fig:EP_Hist_pyramid}, the date-of-first-monitoring and the discovery date for the detected exoplanets were 
largely obtained from the literature and press releases (Table 3). 
In some cases the date-of-first-monitoring was estimated from the first point on the planet host's velocity curve.
The ramp-up time needed to start observing all the stars in a survey's target list was
estimated from the distribution of the date-of-first-montitoring of a survey's detected exoplanets.
That is, the ramp up time needed to start observing the detected hosts was used to estimate
the ramp-up time needed to start observing all the stars on a survey's target list.
The exceptions to this were the time dependence of the original Lick  
target list obtained from Cumming \etal (1999) 
and the time dependence of the Keck target list (FGK) estimated from 
the FGKM histogram of Cumming \etal (2003).  

The date-of-first-monitoring binning (middle-gray histogram in the top panel of Fig. \ref{fig:EP_Hist_pyramid}) shows that 
the fraction of target stars hosting planets has a maximum of $\approx 11\%$ from the longest monitored targets and then 
decreases as we average in target stars that have been monitored for shorter times.
The  two binning conventions used (date-of-detection and date-of-first-monitoring) are equivalent today at
the current value of $5\% (\approx 92/1812)$. 

\begin{figure}
\epsscale{0.8}
\plotone{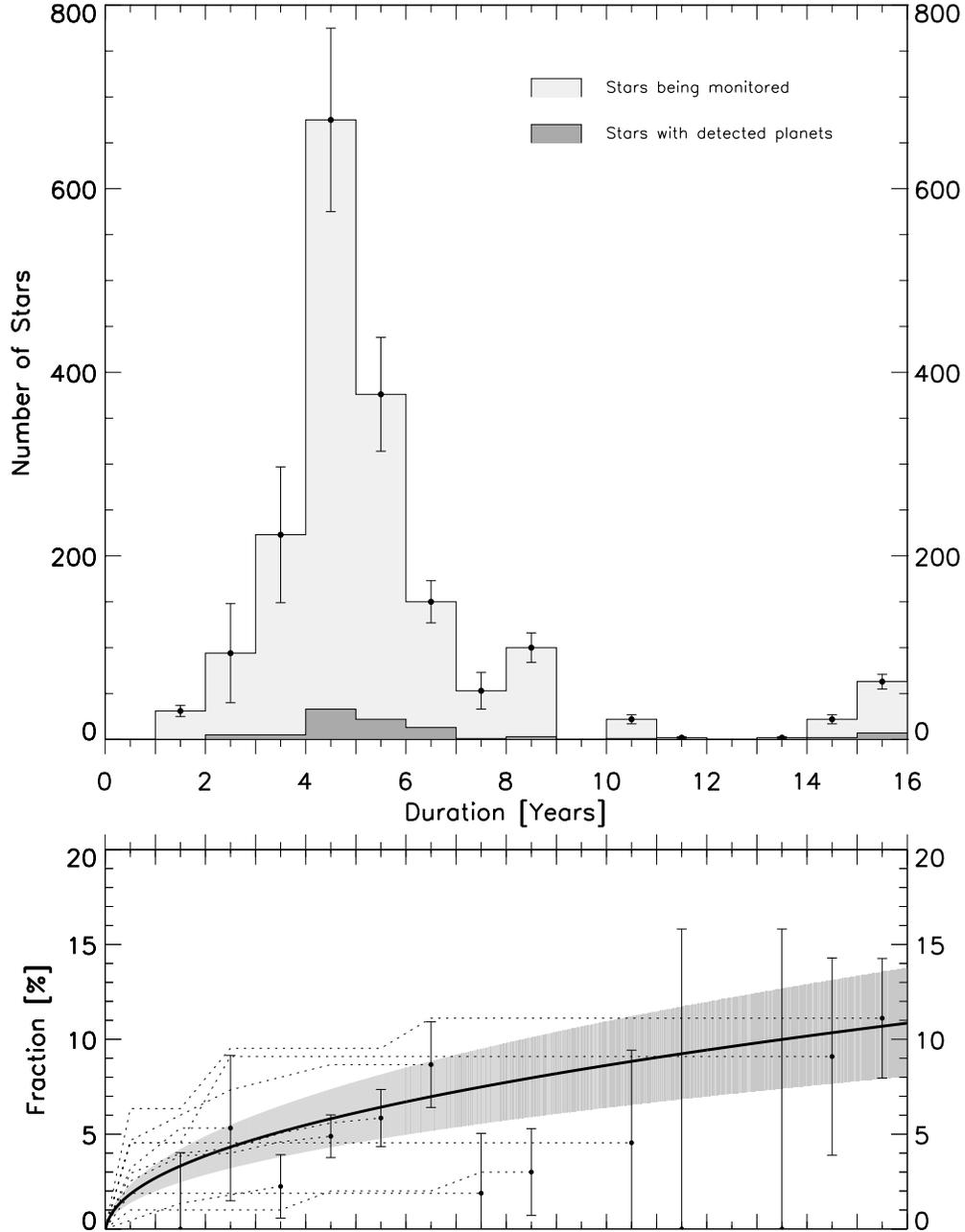}
\caption{\scriptsize
Number of targets as a function of how many years they have been monitored.
Most targets have only been monitored for $\sim 5$ years.
The longer a group of target stars has been monitored, the larger the fraction of planet-possessing targets.
TOP: 
Notice the small number of target stars ($\sim 90$) in the original Doppler surveys (bins on far right) and 
the $\sim 2$ years it took to start monitoring all of them. 
These $\sim 90$ ($\sim 5 \%$  of the stars currently being monitored) are the only target 
stars that have been monitored long enough to begin to detect exoplanets with Jupiter-like masses in Jupiter-like ($\sim 12$ 
year) orbital periods. This sample in the two longest duration bins has resulted in 
9 detections from 85 targets for a fraction of $\approx 11 \%$. 
It will be another five years before a substantially larger fraction of target stars have been monitored long enough 
to detect such planets.
BOTTOM:
The fractions plotted here are the ratios of the histograms in the upper panel as in the previous figure.
This panel shows the effect of longer monitoring duration on the observed fraction of stars with planets. 
The fraction of target 
stars with detected planets increases with duration. This is distinct from the cumulative fractions 
shown in the lower panel of Fig. \ref{fig:EP_Hist_pyramid}. However, the right-most bin here is the 
same as the left-most bin there.  Calculation of the fraction of stars hosting exoplanets
(the curve in the lower panel with the light gray 68\% confidence region) is described in Sec. \ref{sec:KPplane}.
%
The scatter of the data points around the smooth curve is due to small number statistics, varying instrumental 
sensitivities and variation in observational phase coverage. For example, the two points at durations of 7.5 and 8.5 years 
lie  below the curve, due to the lower average sensitivity $\gtrsim 10$ m/s of AFOE and Elodie surveys in these bins.
The high point at 6.5 years is the start of the sensitive Keck survey. 
Thus, this plot also shows 
the effect of different instrumental sensitivities on the fraction of stars hosting planets. 
}
\label{fig:EP_Hist_timeline}
\end{figure}

Although this $11\%$ estimate is based on the results from $\sim 85$ 
target stars of the two longest running surveys: Lick and McDonald, 
it is consistent with the increasing fraction based on the monitoring duration of all target stars.
This is shown in Fig. \ref{fig:EP_Hist_timeline}.
We fit a curve to this data normalized at a duration of 15 years to the weighted average fraction of the two longest 
duration bins (Sec. \ref{sec:KPplane}).
The extrapolation of this curve (based on current sensitivity) to monitoring durations of 30 years 
(approximate period of Saturn) yields a fraction of $\sim 15 \%$. This extrapolation corresponds to extending 
the white ``Detected'' region in Fig. \ref{fig:EP_m_p_big} to the upper right, above the diagonal $K = 40$ m/s line.
%

In the lower panel of Fig. \ref{fig:EP_Hist_timeline}, we increase the fraction for a given group of stars, 
not when a planet is reported, but at the duration equal to the period of the newly detected planet. For example, recent 
detections of short period planets increase the fraction at a duration 
corresponding to the short period of the newly detected planet. This smooths over artificial delays associated, 
for example, with not analyzing the data for the first 8 years of observations, and enables us to trace with dotted lines
the increasing fraction for each group of target stars.

\clearpage

\clearpage

\clearpage
\begin{figure}[!pht]
\epsscale{0.8}
\plotone{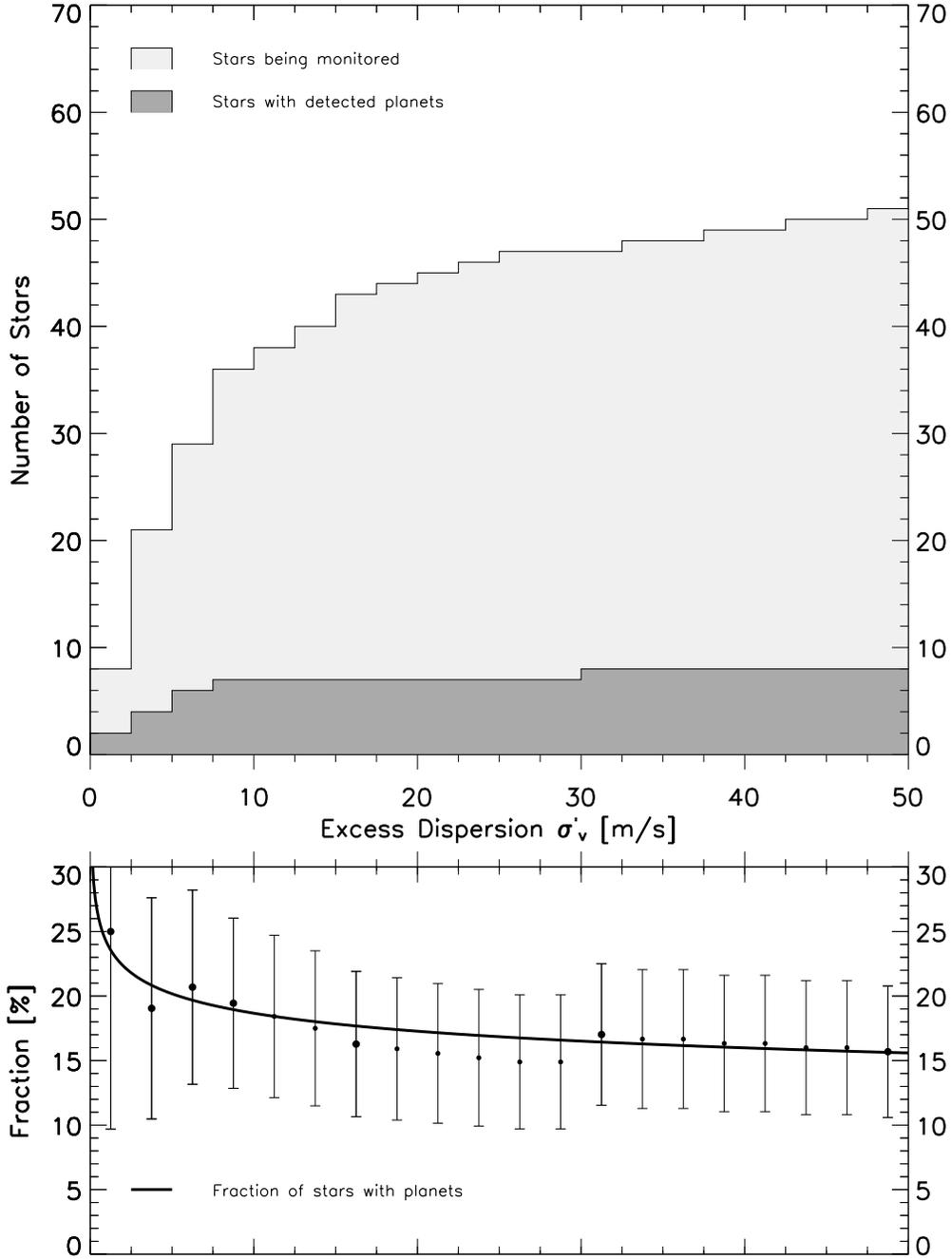}
\caption{\scriptsize
When stellar activity is low, high measurement precision is possible and a higher fraction
of targets are found to host planets.
TOP:
The number of targets in the original Lick survey that have been monitored for a duration of 14-16 years
and that have an excess radial velocity dispersion $\sigma'_{v}$ less than the value on the x axis.
$\sigma'_{v}$ is a measure of how precisely one can determine the radial velocity of the targets.
The  number of detections from these targets is also shown. 
The ratio (BOTTOM) gives the fraction of stars with planets as a function
of $\sigma'_{v}$ threshold.
The $\sigma'_{v}$ of the Lick target stars have been estimated 
from the stellar rotational periods tabulated by Cumming \etal (1999) using the relations
$\sigma'_{v} = 10*(12/P_{rot})^{1.1}$ for G and K stars
and $\sigma'_{v} = 10*(10/P_{rot})^{1.3}$ for F stars 
(Saar \etal 1998).
%
%
The curve is a power law ($df/d\sigma'_{v} \propto \sigma'_{v}{}^{\gamma}$)
fit to a sparse sample of these non-independent points (larger point size).
Thus, as we select for higher Doppler precision in the Lick sample, the fraction of targets possessing planets
increases from $15\%$ to $25\%$ for the stars with the highest precision
($\sigma'_{v} < 2.5$ m/s). Small number statistics increases the error bars as the decreasing $\sigma'_{v}$
threshold reduces the number of targets. In this cumulative plot the error bars are highly correlated. 
}
\label{fig:highDoppler}
\end{figure}

\clearpage
\section{High Doppler Precision Targets}
\label{sec:highDopp}

Instrumental sensitivity is an important limit on a Doppler survey's ability to detect planets. 
However, the level of stellar activity on a star's surface is also important. 
Using the Doppler technique, planets are more easily detected around slowly rotating stars with low level chromospheric 
activity, little granulation or convective inhomogeneities and few time-dependent surface features.  
To reduce these problems, some target lists have been selected for high Doppler precision (low stellar activity)
by excluding targets with high values of  projected rotational velocity $\vsin$ or chromospheric emission 
ratio $R'_{HK}$ (Table 3).

By selecting target stars monitored the longest, the detected fraction of Sun-like stars possessing planets
increased from $5\%$ to $11\%$. By selecting from the target stars monitored the longest, the stars with the highest
Doppler precision, the fraction increases still further. 
Fischer \etal (2003) analyzed a group of high Doppler precision target stars 
in the original Lick sample and found that $15\%$ possessed planets. 
We extend this idea in Fig. \ref{fig:highDoppler} by plotting the fraction of target stars with planets as a function of 
their stellar activity as measured by their excess radial velocity dispersion 
$\sigma'_{v}$ (Saar \etal 1998).
We find that this selection increases the fraction from $15\%$ on the far right of Fig. \ref{fig:highDoppler}
to $\sim 25 \%$ on the far left for the highest Doppler precision stars.
As long as high precision Doppler stars are an unbiased sample of Sun-like stars
this indicates that at least $\sim 25\%$ of Sun-like stars possess planets.

However, as we decrease the $\sigma'_{v}$ threshold to consider only the highest precision stars, 
we are using fewer target stars to infer the fraction. Thus, the error bars on the estimates increase
from $15 \pm 5\%$ on the right to $25 \pm 15\%$ on the left.
\section{Fitting for and extrapolating trends}
\label{sec:slopes}
\begin{figure} 
\epsscale{0.9}
\plotone{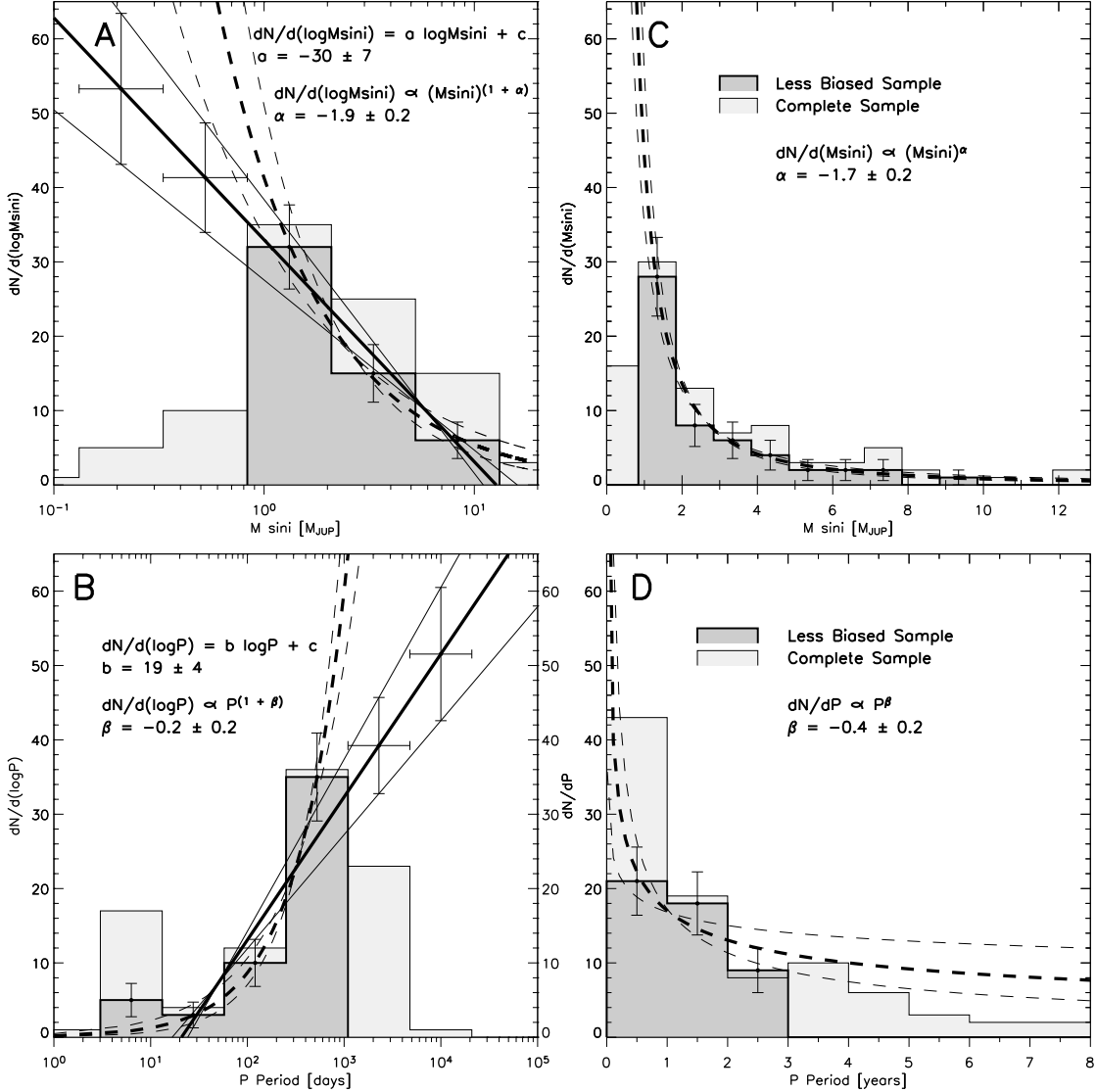}
\caption{\scriptsize 
Trends that we extrapolate to lower mass and longer periods.
Histograms of planet mass (top) and period (bottom) for the less-biased sample of exoplanets within 
the thin solid rectangle of Fig. \ref{fig:EP_m_p_big}, compared to all exoplanet detections.
Both log (left) and linear (right) versions are shown.
Histograms in log $\msin$ and log $P$ (panels A \& B) are fitted with the linear functional forms 
$dN/d(logMsini) = a\;logMsini + c$ and 
$dN/d(logP) = b\;logP + d$ where the best fits to the histograms yield $a = -30 \pm 7$, 
and $b = 19 \pm 4$.
The linearly binned histograms of $\msin$ and $P$ (panels C \& D) 
are fitted to the functional forms: $dN/d(\msin) \propto (\msin)^{\alpha}$ and $dN/dP \propto P^{\beta}$ respectively 
and we find $\alpha = -1.7 \pm 0.2$ and $\beta = -0.4 \pm 0.2$. 
To check the robustness of the fits we change variables to log $\msin$ and log $P$ (effectively producing a re-binning
of the data) and fit for $\alpha$ and $\beta$ in panels A \& B.
We find $\alpha = -1.9 \pm 0.2$ and $\beta = -0.2 \pm 0.2$. 
Combining these two estimates gives our best estimates of $\alpha = -1.8 \pm 0.3$ and $\beta = -0.3^{+0.3}_{-0.4}$. 
In the fit of panel B, we ignore the 5 exoplanets in the smallest period bin because we are interested in the overall pattern 
that can be most plausibly extrapolated to larger $P$ bins, not in the pile up associated with a 
poorly understood stopping mechanism at $P < 12$ days.
For consistency, these 5 exoplanets have also been removed from the smallest period bin in panel D.
}
\label{fig:EP_Hist_log_lin}
\end{figure}

We use extrapolation to estimate the fraction of stars with planets
within regions of mass-period parameter space larger than the less-biased sample (Fig. \ref{fig:EP_m_p_big}).
Since there is no accepted theoretical model for the functional form
for the mass or period distribution functions, we make simple linear fits to the
histograms in log $\msin$ and log $P$ and we fit conventional power laws to the
histograms in $\msin$ and $P$ (Fig. \ref{fig:EP_Hist_log_lin}).

\clearpage
\clearpage
%
Fitting $dN/d(logMsini) = a \; logMsini + c$ to the histogram of log $\msin$ (Fig. \ref{fig:EP_Hist_log_lin}{\bf A})  
we obtain the slope $a= -30 \pm 7$.
Fitting $dN/d(logP) = b \; logP + d$ to the histogram of log $P$ (Fig. \ref{fig:EP_Hist_log_lin}{\bf B})
we obtain the slope $b = 19 \pm 4$.   
These trends are shown as thick lines.  
The distribution of exoplanets is not flat in either log $\msin$ or log $P$. That is $a = -30 \pm 7$ is 
significantly different ($\sim 4 \sigma$) from $a = 0$ and $b = 19 \pm 4$ is significantly 
different ($\sim 5  \sigma$) from $b = 0$.
Fitting $dN/d(Msini) \propto (Msini)^{\alpha}$ to the histogram of  $\msin$ 
(Fig. \ref{fig:EP_Hist_log_lin}{\bf C}) we obtain $\alpha = -1.7 \pm 0.2$. 
Fitting $dN/dP \propto P^{\beta}$ to the histogram of $P$ (Fig. \ref{fig:EP_Hist_log_lin}{\bf D})
we obtain $\beta = -0.4 \pm 0.2$.

Since $d\; (ln\;x) = dx/x$, the functional form $dN/d(Msini) \propto (Msini)^{\alpha}$ can be written 
as $dN/d(log \msin) \propto (\msin)^{(1 + \alpha)}$ and similarly $dN/dP \propto P^{\beta}$ can be written as
$dN/d(logP) \propto P^{(1 + \beta)}$. We fit these functions to the histograms of log $\msin$ and 
log $P$ in Fig. \ref{fig:EP_Hist_log_lin}{\bf A} \& {\bf B}
where we find $\alpha = -1.9 \pm 0.2$ and $\beta = -0.2 \pm 0.2$. 
These values differ by $\sim 1 \sigma$ from the values of $\alpha$ and
$\beta$ found in Fig. \ref{fig:EP_Hist_log_lin}{\bf C} \& {\bf D}.
We attribute the differences to the different log and linear binning schemes.
We combine the two values into our best estimates:
$\alpha = -1.8 \pm 0.3$ and $\beta = -0.3^{+0.3}_{-0.4}$. 
These values are compared to other estimates in Table 1.   
%
If the distributions were flat in log $\msin$ and log $P$ 
we would find respectively $\alpha \approx -1.0$ and $\beta \approx -1.0$. However, both
the trend in mass ($\alpha$) and in period ($\beta$) 
are significantly different ($\sim 2 \sigma$) from flat.
These slopes agree with our previous results (Table 1) 
%
and show that the evidence supporting the idea that 
Jupiter 
lies in a region of parameter space densely occupied by exoplanets,
has gotten stronger in the sense that
our new value $b=19$ is larger than our previous estimate and our new values for $a$ and
$\alpha$ are more negative than our previous estimates. 

The main differences between our results and previous results is that we obtain a steeper slope 
(more negative $\alpha$) when we fit the funtional form $dN/d(\msin) \propto  (\msin)^{\alpha}$ to 
the mass histogram of the data: we get $\alpha \approx -1.8$ while other analyses get $\alpha \approx -1.1$. 
Thus we predict more low mass
planets relative to the more easily detected number of high mass planets, than do other analyses.
We obtain a more shallow slope (less negative $\beta$) when we fit the funtional form $dN/dP \propto  P^{\beta}$ to 
the period histogram of the data.
We get $\beta \approx -0.3$ while other analyses get $\beta \sim -0.8$. Thus, we predict more hosts of long period
planets relative to the more easily detected short  period planets, than do other analyses.
These differences are largely due to the differences in the way the incompleteness in the lowest mass bin and the
longest period bin are accounted for. This can be seen, for example in the $\alpha = -0.7$ reported by 
Marcy \etal (2003) when no correction is made for incompleteness in the lowest mass bin.
When we make no completeness correction and include the lowest mass bin, we reproduce their result.

\begin{figure}   
\vspace{1cm}
\epsscale{0.4}
\plotone{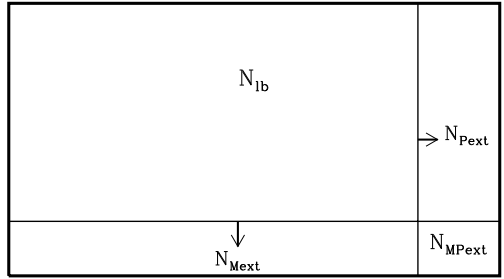}
\vspace{1cm}
\caption{The simple method of extrapolation used to predict planet numbers in under-sampled
regions of the log mass - log period plane. 
The known value of $N_{lb}$ (number of planet hosts in the less-biased region of Fig. \ref{fig:EP_m_p_big})
and extrapolations based on the slopes $a$ and $b$ were used to derive
$N_{Mext}$ and $N_{Pext}$. We then use Eq. \ref{eq:ratios}
to estimate $N_{MPext}$.
Our estimate of the total number of planet hosts is then $N_{hosts} = N_{lb} + N_{Mext} + N_{Pext} + N_{MPext}$.
}
\label{fig:boxes}
\end{figure}

\subsection{Extrapolation Using Discrete Bins}
\label{sec:discrete}
Based on slopes $a$ and $b$, 
we make predictions for the fraction of Sun-like stars with planets within two regions of the log mass - log period plane.
The estimated populations of the two lowest mass bins based on the slope $a$ are indicated in 
Fig. \ref{fig:EP_Hist_log_lin}{\bf A}. Similarly the estimated populations of the two largest 
period bins based on slope $b$ are shown in Fig. \ref{fig:EP_Hist_log_lin}{\bf B}. 
From the less-biased region (thin rectangle in Fig. \ref{fig:EP_m_p_big}) we extrapolate both in $\msin$ and $P$ by 1 bin.  
When we extrapolate in $\msin$ or $P$ we are doing so at a fixed range in $P$ or $\msin$ respectively.
However to estimate the fraction of stars within 
the thick rectangle in Fig. \ref{fig:EP_m_p_big} (given by the $\msin$ and $P$ ranges $ 0.3 < \msin/M_{Jupiter} < 13$ 
and $ 3\; \mbox{\rm days} < P < 13\; \mbox{\rm years}$) we need to estimate 
the number of planets $N_{MPext}$ in the bottom right-hand region (just below Jupiter, see Fig.
\ref{fig:EP_m_p_big}).
To do this we use Eq. \ref{eq:ratios} (see Fig. \ref{fig:boxes}).
We have  $N_{lb} = 53 \pm 2$      
from Fig. \ref{fig:EP_m_p_big} and
$N_{Mext} = 41 \pm 7$ and $N_{Pext} = 39 \pm 6$ from Fig. \ref{fig:EP_Hist_log_lin}{\bf A}\&{\bf B}.
Inserting these values into, 
\begin{eqnarray}
\frac{N_{Mext}}{N_{lb}} \approx \frac{N_{MPext}}{N_{Pext}},
\label{eq:ratios}
\end{eqnarray}
(see Fig. \ref{fig:boxes}) we obtain $N_{MPext} = 30 \pm 9$. Thus, 
$N_{hosts} = 163 \pm 24$ and
the fraction $f$, of targets hosting planets is the ratio,
\begin{eqnarray}
f &=& \frac{N_{hosts}}{N_{targets}},
\label{eq:lbs}
\end{eqnarray}
where $N_{targets} \sim 1812 \pm 103$ (Table 4). Finally we find 
$f = (163 \pm 24)/ 1812 \pm 103) \approx 9 \pm 1 \%$.
Thus, using the slopes $a$ and $b$ to extrapolate one bin into lower 
masses and longer periods (thick solid rectangle in Fig. \ref{fig:EP_m_p_big}) 
we find that $9 \pm 1 \%$ of the targeted Sun-like stars have planets.

Similarly and more speculatively we estimate the fraction contained within 
the thick dashed rectangle in Fig. \ref{fig:EP_m_p_big} by extrapolating one more bin in mass (over the same range in
period as the previous mass extrapolation) and by extrapolating one more bin in period (over the same range in 
mass as the previous period extrapolation). The analogous numbers are
$N_{lb} = 53 \pm 2$, $N_{Mext} = (41 \pm 7) + (53 \pm 10)$, $N_{Pext} = (39 \pm 6) + (52 \pm 9)$ 
(where the sum of the 2 sets of numbers is the sum of the 2 separate bins). Equation \ref{eq:ratios}
then yields $N_{MPext} = 161 \pm 50$. Summing the four contributions as before  yields
$N_{hosts} = 399 \pm 84$ and thus $f = (399 \pm 84)/ 1812 \pm 103) \approx 22 \pm 5 \%$.
Thus, we estimate that $22 \pm 5 \%$ 
of the monitored Sun-like stars have planets in the larger region ($ 0.1 < \msin /M_{Jupiter} < 13$ and 
$ 3\; \mbox{\rm days} < P < 60\; \mbox{\rm years}$) that encompasses both Jupiter and Saturn. 
This larger region is less than $20\%$ of the area of the log mass - log period plane occupied by our planetary system.    

\begin{table*}  
\begin{center}
\caption{Best-fit trends in fits to mass and period histograms and comparison with other analyses}
\begin{tabular}{|l|c|c|c|c|}
\hline
Source                                     & $\alpha$         & $\beta$          & $a$         & $b$    \\
\hline
this paper,  Fig. 4 A,B$^{j}$              &     --           &    --            & $-30 \pm 7$ & $19 \pm 4$ \\  
``    ''  Fig. 4 C,D$^{k}$                 & $-1.7 \pm 0.2 $  & $-0.4 \pm 0.2$   &    --       &     --     \\
``     '' Fig. 4 A,B$^{l}$                 & $-1.9 \pm 0.2$   & $-0.2 \pm 0.2$   &    --       &    --     \\  
``     ''combined Fig. 4 A,B,C,D           & $-1.8 \pm 0.3$   &$-0.3^{+0.3}_{-0.4}$ &    --       &  --        \\
Marcy, Butler, Fischer \& Vogt 2003        & $-0.7^{m}$       &        --        &    --       &     --     \\
Lineweaver, Grether \& Hidas 2003          & $-1.6 \pm 0.2$   &     --           &    --       & $13 \pm 4$ \\
Lineweaver \& Grether 2002                 & $-1.5 \pm 0.2$   &     --           & $-24 \pm 4$ & $12 \pm 3$ \\
Tabachnik \& Tremaine 2002                 & $-1.11 \pm 0.10$ & $-0.73 \pm 0.06$ &    --       &   --       \\
Stepinski \& Black 2001                    & $-1.15 \pm 0.01$ & $-0.98 \pm 0.01$ &      --     &    --      \\
Jorissen, Mayor \& Udry 2001               & $\sim \;-1$      &      --          &      --     &   --       \\ 
Zucker \& Mazeh 2001                       & $\sim \;-1$      &        --        &      --     &    --      \\
\tableline 
\end{tabular}\\
\end{center}
\scriptsize
\noindent

$^{j}$ best-fit of $dN/d(log \msin) = a \; log \msin + c$ to histogram of log $\msin$ and
       best-fit of $dN/d(logP) = b \; logP + d$ to histogram of log $P$\\
$^{k}$  best-fit of $dN/d(\msin) \propto (\msin)^{\alpha}$ to histogram of $\msin$ and
       best-fit of $dN/dP \propto P^{\beta}$ to histogram of $P$\\ 
$^{l}$ best-fit of $dN/d(log \msin) \propto (\msin)^{1+\alpha}$ to histogram of log $\msin$ and
       best-fit of $dN/d(log P) \propto P^{1+\beta}$ to histogram of log $P$\\
$^{m}$ including the lowest mass bin with no completeness corrections\\
\\
\label{table:slopes}
\end{table*}

\subsection{Extrapolation Using a Differential Method}
\label{sec:diff}
Instead of extrapolating one or two discrete bins, we can generalize to a more flexible differential method.
For example, we can use the power law functional form to integrate a differential fraction
within an arbitrary range of $\msin$ and $P$ \citep{Tabachnik02},

\begin{equation}
df = c (\msin)^{\alpha} P^{\beta} \; d(\msin) dP. 
\label{eq:TT}
\end{equation}
%
%
We find the normalization $c$ by inserting the known values from the less-biased area:
$f = (53 \pm 2/(1812 \pm 103)) = 2.9 \pm 0.2\%$,
$\alpha = -1.8 \pm 0.3$ and $\beta = -0.3^{+0.3}_{-0.4}$ (Table 1).  
We integrate between the boundaries of the less-biased rectangle 
($3$ days $ < P < 3$ years and $0.84 < \frac{\msin}{M_{Jupiter}} < 13$) and solve for the normalization. 
We find $c = 8.6^{+40.7}_{-5.8} \times 10^{-5}$.
Under the assumption that this same $c$, $\alpha$ and $\beta$ hold over larger regions of parameter space, 
we can integrate over the larger region and solve for $f$.
For the thick solid and thick dashed regions shown in Fig. \ref{fig:EP_m_p_big}, we find 
$f = 20^{+23}_{-10} \%$ and $100^{+0}_{-69} \%$ respectively. 
Using this differential method with the slopes $a$ and $b$ we find
values nearly identical with the discrete bin extrapolations based on $a$ and $b$: 
$f = 9 \pm 1 \%$ and $22 \pm 5 \%$ respectively. 

The power-law based estimates are consistent with the lower linear ($a-$ and $b-$based) estimates
in the sense that the large error bars on the power-law estimates overlap with the lower estimates of the linear method.
The reason the power law fits yield larger fractions can be seen in Fig \ref{fig:EP_Hist_log_lin} {\bf A} \&{\bf B}.
In the low mass and large period regions, the dashed curves are higher than the solid lines.
Without further data from less massive planets and larger periods we interpret this difference as an
uncertainty associated with the inability of the data to prefer one of the two simple functional forms fit here.
The power law fits are marginally better fits to the data.
We interpret the lower values from the linear fits as conservative lower limits to the fraction $f$.

\subsection{Fractions in Velocity-Period Parameter Space}
\label{sec:KPplane}

The fractions discussed in Sections \ref{sec:exoplanetdata} \& \ref{sec:highDopp} 
are based on exoplanet detections constrained by the Doppler technique to a 
trapezoidal region of the log mass - log period plane defined by a minimal
velocity $K_{min}$. The fractions discussed in the previous two sections are based on rectangular regions of 
the log mass - log period plane.
In the lower panel of Fig. \ref{fig:EP_Hist_timeline}, the fractions plotted as a function of duration come from
a trapezoidal region (not a rectangular region) of the $\msin - P$ plane.
Hence, a fit of $dN/dP \propto P^{\beta^{\prime}}$ to the $P$ histogram of 
detected planet hosts, will produce a value of $\beta^{\prime}$ slightly different from the $\beta$ of  
Fig. \ref{fig:EP_Hist_log_lin}{\bf D}.
We find $\beta^{\prime}= -0.5 \pm 0.2$ (while $\beta = -0.4 \pm 0.2$). Such a difference is expected since the
number of planets at large $P$ will decrease more steeply in the trapezoidal region, because as $P$ increases the
``Detected'' regions becomes narrower than a region defined by a constant $\msin$.
The curve shown in the lower panel of Fig. \ref{fig:EP_Hist_timeline} is the integral of $P^{\beta^{\prime}}$.
It is normalized to the last two bins on the right.
In plotting this integral we are assuming that a survey of duration $P_{s}$ has observed a large fraction of its target
stars with Doppler-detectable exoplanets of period $P \lesssim P_{s}$.

An interesting consistency check of the relationships between our extracted values of $\alpha$, $\beta$ and $K_{min}$
and the limits of integration for Eq. \ref{eq:TT} is described in Appendix B.

\subsection{Jupiter-like Planets}
\label{sec:jupiters}

The fraction of Sun-like stars possessing Jupiter-like planets is important since Jupiter is the dominant orbiting 
body in our Solar System and had the most influence on how our planetary system formed. 
Exoplanets with Jupiter-like periods and Jupiter-like masses are on the edge of the detectable region of parameter space. 
The cross-hatched square in Fig. \ref{fig:EP_m_p_big} is our Jupiter-like region defined by
orbital periods between the period of the asteroid belt and Saturn
with masses in the range $M_{Saturn} < \msin < 3 M_{Jupiter}$.
We estimate the fraction of Sun-like stars hosting planets in this region using the differential method
based on our best-fit values of $a$ and $b$.
Integrating between the limits of the cross-hatched area we obtain $f = 5 \pm 2\%$
of Sun-like stars possess such a Jupiter-like planet.
Using the differential method based on our best-fit values for $\alpha$ and $\beta$
we obtain  $f = 28^{+62}_{-21} \%$. 
As with the different $f$ values resulting from these two methods in Sec. \ref{sec:diff},
these are consistent with each other and  reflect the different functional forms used.


\begin{figure}
\epsscale{0.9}
\plotone{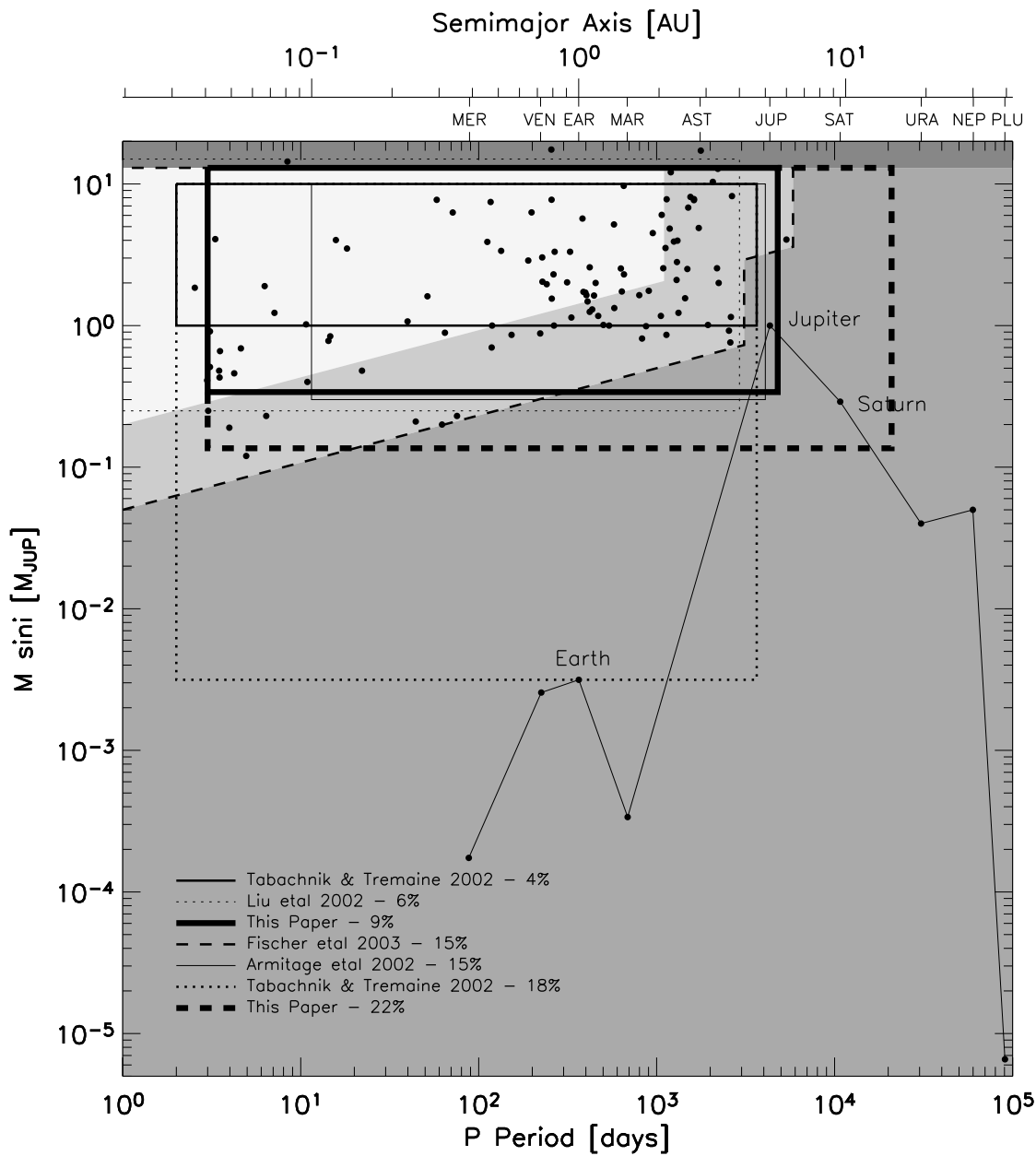}
\caption{We estimate that at least $9 \%$ of Sun-like stars have planets within the thick solid rectangle and at
least $22\%$ have planets within the thick dashed rectangle (Sec. \ref{sec:discrete}).
We find larger values using a differential method based on power-law fits (Sec. \ref{sec:diff}).
Here, we compare our results with other published estimates of the fraction of stars with planets.
The papers, estimates and the regions of the log $\msin$ - log $P$ plane associated with these estimates
are indicated.
Our results are consistent with, but generally higher than, previous work (Table 2).
}
\label{fig:comparison}
\end{figure}

\begin{table*}
\begin{center}
\caption{Fraction Comparison (see Fig. \ref{fig:comparison})}
\small
\begin{tabular}{|l|c|c|c|c|c|}
\hline
Source                     & $\msin$ range     & Period range       & Fraction & Our Fraction$^{a}$ & Our Fraction$^{b}$ \\
\hline

TT$^{c}$ 2002            & 1 - 10 $M_{Jup}$    & 2 days - 10 yrs      & 4\%   & $4 \pm 1\%$       & $6^{+3}_{-2}\%$  \\
Liu \etal 2002           & 0.25-15 $M_{Jup}$   & 0-8 yrs ($<$ 4 AU)   & 6\%   & $9 \pm 3\%$       & $18 ^{+19}_{-7} \%$  \\
Armitage \etal 2002      & 0.3-10 $M_{Jup}$    & 0.03-11.2 yrs$^{d}$  & 15\%  & $9 \pm 3\%$       & $19 ^{+22}_{-10} \%$  \\
TT$^{c}$ 2002            & 0.003-10 $M_{Jup}$  & 2 days - 10 yrs      & 18\%  & $45^{+19}_{-15}\%$& $100 \%$ \\

\tableline 
\end{tabular}
\end{center}
\scriptsize
\noindent
$^{a}$  differential method based on slopes $a$ and $b$\\
$^{b}$  differential method based on powers $\alpha$ and $\beta$\\
$^{c}$ Tabachnik \& Tremaine (2002)\\  
$^{d}$ corresponds to 0.1 - 5 AU\\
\end{table*}

\section{Summary and Discussion}
\label{sec:summary}

We have analyzed the results of eight Doppler surveys to help answer the question:
`What fraction of stars have planets?'. 
We use the number of targets and the number of detected planet hosts to
estimate the fraction of stars with planets.
Quantitatively following the consistent increase of this fraction is an important new way to 
substantiate both current and future estimates of this fraction. 
We show how the naive fraction of $\sim 5\%$ increases to $\sim 11\%$ when only long-duration targets
are included.
We extend the work of Fischer \etal (2003) by plotting the fraction as a function of
excess dispersion and show how this $\sim 11\%$ increases to $\sim 25\%$ when only long-duration Lick 
targets with the lowest excess in radial velocity dispersion are considered. 

We have identified trends in the exoplanet data based on a less-biased sub-sample.
We find stronger support than found previously
for the idea that Jupiter 
-like planets are common in planetary systems
(Table 1).
We estimate the fraction of Sun-like stars hosting planets in a well-defined Jupiter-like
region to be $\sim 5 \%$.   

We have extrapolated these trends into unsampled or undersampled regions of the log mass - log period plane.
We find at least $\sim 9 \%$ 
of target stars  will be found to host an exoplanet within the thick rectangle of Fig. \ref{fig:EP_m_p_big}
and that more speculatively at least  $\sim 22 \%$    
of target stars will be found to host an exoplanet within the larger thick dashed 
rectangle. 
Our results for the fraction of Sun-like stars with planets
are consistent with but are, in general, larger than previous estimates (Table 2, Fig.~\ref{fig:comparison}). 

The largest uncertainty in this analysis is that we may be extrapolating trends
derived from a small region of log mass - log period into regions of parameter space in which the 
trends are slightly or substantially different. This uncertainty is why we did not extrapolate beyond
the dashed region in Fig. \ref{fig:EP_m_p_big} that contains from our Solar System, only Jupiter and Saturn.

It is sometimes implicitly assumed that most planetary systems will be like ours and that Earth-like planets
will be common in the Universe.
However, as we descend in scale from galaxy, to star, to planetary system to terrestrial planet
we run more of a risk of self-selection. That is, the factors that are responsible for 
our origin may have selected a non-typical location. 
Thus, answering the question ``What fraction of stars have planets?'' must rely on the continued analysis of the
statistical distributions of exoplanets detected by the increasingly precise and ground-breaking Doppler surveys.

The hypothesis that $\sim 100\%$ of stars have planets is consistent with both the observed exoplanet data which 
probes only the high-mass, close-orbitting exoplanets and with the observed frequency of circumstellar
disks in both single and binary stars. The observed fractions $f$ that we have derived from current exoplanet data
are lower limits that are consistent with a true fraction  of stars with planets $f_{t}$, in the 
range $0.25 \lesssim f_{t} \lesssim 1$.
If the fraction of Sun-like stars that possess planets is representative
of all stars, our result means that out of the $\sim 300$ billion stars in our Galaxy there are
between $\sim 75$ and $\sim 300$ billion planetary systems.


\section{Acknowledgements}
CHL acknowledges a research fellowship from the Australian Research Council.
We thank Stephane Udry, Debra Fischer, Andrew Cumming  and Sylvain Korzennik for help with many details
of the exoplanet surveys. We thank Luis Tenorio for statistical counseling.



\begin{table}   
\label{table:surveys}
\begin{center}
\caption{Doppler Surveys$^{a}$ - Targets}
\tiny
\begin{tabular}{|l|l|l|l|l|l|l|l|l|} \hline
                             & AAT & Lick     & Keck          & Coralie      & Elodie  & AFOE    & CES        & McDonald   \\
\hline
Targets$^{b}$                & 204  & 360     & 600$^{c}$     & 1100$^{d}$   & 350     & 146     & 37         & 33         \\
FGK IV, V Targets            & 198  & 360     & 443           & 1100         & 350     & 136     & 32         & 33         \\
List Published               & Y    & Y$^{e}$ & Y$^{e}$       & N            & N       & Y$^{f}$ & Y          & Y          \\
Year Started                 & 1998 & 1987    & 1996          & 1998         & 1994    & 1995    & 1992       & 1987       \\
Spectral Types               & FGKM & F7-K0   & F7-M5         &F8-M0         & FGK     & FGK     & F8-M5      & FGK        \\
Apparent V                   & $\lesssim 7.5$ & $\lesssim 7.5$& $\lesssim 11$& $<$9 (for 80\%) & $<$7.65 & $\lesssim 7.5$ 
                             & $\lesssim 6$   & $\lesssim 6$ \\ 
Log($R'_{HK}$)$^{g}$         & $\lesssim -4.5$&$\lesssim -4.5$& $\lesssim -4.5$&    &   &  &    &    \\
vsini (km/s)$^{h}$                 &      &         &               &$<$4$^{d}$    & $<$ 5   & $<$ 8   &            & \\
Detected SB1's \% $^{i}$     & 8.8  & 3.3     & 3.3           & 8.7          & 3.5     & 3.5     & 8.1        & 3.5      \\ 
Planet Host Detections$^{j}$ & 18   & 18      & 26            & 33           & 16      & 6       & 2          & 3       \\
References$^{k}$             & 1    & 2,9,10  & 3,9,11        & 4            & 5,12    & 6       & 7          & 8       \\
\tableline 
\end{tabular}\\
\end{center}
\scriptsize
\noindent
$^{a}$ AAT - Anglo Australian Observatory, Anglo Australian Telescope, UCLES Spectrograph \\
Lick - Lick Observatory, Hamilton Spectrograph \\ 
Keck - Keck Observatory, HIRES Spectrograph \\
Coralie - European Southern Observatory, Euler Swiss Telescope, Coralie Spectrograph \\
Elodie - Haute Provence Observatory, Elodie Spectrograph \\
AFOE - Whipple Observatory, AFOE Spectrograph \\
CES - European Southern Observatory, CAT Telescope, CES Spectrograph \\
McDonald - McDonald Observatory, Coud\'{e} Spectrograph \\
$^{b}$ refers to all target stars in a survey \\
$^{c}$ another 450 stars have one or more observations but have been dropped for various reasons (Butler \etal 2002) \\
$^{d}$ there are 550 faster rotators in a lower priority target list (Udry 2003, private communication)\\
$^{e}$ combined Lick/Keck (889 stars)\\
$^{f}$ Korzennik 2003 private communication \\
$^{g}$ the fractional CaII H and K flux corrected for the photospheric flux (see Noyes 1984 and Saar \etal 1998)\\
$^{h}$ projected rotational velocity\\
$^{i}$ SB1: single-lined spectroscopic binaries.The AAT survey \citep{Jones02a} finds 18 out of 204 target stars are SB1.
\citep{Nidever02} mentions that 29 out of 889 Lick and Keck target stars are SB1.
\citep{Endl02} finds 3 out of 37. The original Lick survey has 5 out of 74 \citep{Cumming99}. 
We estimate the fraction of SB1's for the other surveys by noting a survey as either northern or 
southern hemipshere and taking an average of the known surveys in that hemisphere to estimate the fraction of SB1's.
SB2's have been eliminated from the target lists. All surveys exclude binary stars when the angular separation
is less than $2^{''}$.\\
$^{j}$ including confirmations but excluding confirmations of hosts that were known to have a planet prior to the start 
of observation. Thus several hosts of Lick and Elodie that are monitored by Keck and Coralie to increase phase 
coverage are only included in the Lick and Elodie numbers.\\
$^{k}$ 1) Jones \etal 2002
2) Fischer \etal 2003
3) Butler \etal 2002  
4) Udry \etal 2000. The six newest hosts discovered by Coralie and included here were announced during the XIXth 
IAP Colloquium ``Extrasolar Planets: Today and Tomorrow'', Paris, June 2003 and are in preparation: Mayor \etal 2003b, Zucker
\etal 2003 and Udry \etal 2003.
5) Sivan \etal 2000
6) Korzennik 2003, private comm. 
7) Endl \etal 2002
8) Cochran \& Hatzes 1993
9) Nidever \etal 2002
10) Fischer \etal 1999
11) Vogt \etal 2000
12) Mayor \etal website 2003a\\

\end{table}

\begin{table}   
\label{table:targets}
\begin{center}
\caption{Doppler Surveys - Cumulative Numbers of FGK IV-V Targets as a Function of Time}
\tiny
\begin{tabular}{|l|r|r|r|r|r|r|r|r|r|r|r|r|r|r|r|r|}
\hline
Survey($\sigma$ m/s)$^{a}$& 1988$^{b}$ & 1989 & 1990 & 1991 & 1992 & 1993 & 1994 & 1995 & 1996 & 1997 & 1998 & 1999 & 2000 & 2001 & 2002 & 2003\\
\hline

Lick $10-15$     &   44 &   56 &   62 &   64 &   66 &   67 &   73 &   -- &   -- &   -- &   -- &   -- &   -- &   -- &   -- &  -- \\
Lick $3-5$       &   -- &   -- &   -- &   -- &   -- &   -- &   -- &   74 &   74 &   74 &  265 &  360 &  360 &  360 &  360 &  360\\
McDonald $10-20$ &   22 &   33 &   33 &   -- &   -- &   -- &   -- &   -- &   -- &   -- &   -- &   -- &   -- &   -- &   -- &   --\\
McDonald $5-10$  &   -- &   -- &   -- &   33 &   33 &   33 &   33 &   33 &   33 &   33 &   33 &   -- &   -- &   -- &   -- &   --\\
McDonald $\sim 3$&   -- &   -- &   -- &   -- &   -- &   -- &   -- &   -- &   -- &   -- &   -- &   33 &   33 &   33 &   33 &   33\\
CES        $8-15$&    0 &    0 &    0 &    0 &    0 &   29 &   29 &   29 &   29 &   32 &   32 &   32 &   32 &   32 &   32 &   32\\
Elodie $\sim 10$ &    0 &    0 &    0 &    0 &    0 &    0 &    0 &  142 &  142 &  159 &  315 &  350 &  350 &  350 &  350 &  350\\
AFOE $\sim 10$   &    0 &    0 &    0 &    0 &    0 &    0 &    0 &   23 &   90 &  113 &  136 &  136 &  136 &  136 &  136 &  136\\
Keck $2-5$       &    0 &    0 &    0 &    0 &    0 &    0 &    0 &    0 &    8 &  179 &  270 &  326 &  360 &  399 &  443 &  443\\
AAT $\sim 3$     &    0 &    0 &    0 &    0 &    0 &    0 &    0 &    0 &    0 &    0 &   66 &  198 &  198 &  198 &  198 &  198\\
Coralie $\sim 3$ &    0 &    0 &    0 &    0 &    0 &    0 &    0 &    0 &    0 &    0 &   33 &  700 & 1000 & 1100 & 1100 & 1100\\
\hline
Cumulative       &   66 &   89 &   95 &   97 &   99 &  129 &  135 &  301 &  376 &  590 & 1150 & 2135 & 2469 & 2608 & 2652 & 2652\\    
no overlap$^{c}$ &   63 &   85 &   87 &   87 &   89 &  111 &  111 &  210 &  263 &  413 &  790 & 1465 & 1688 & 1782 & 1812 & 1812\\
\hline
non-cumulative   &   66 &   23 &    6 &    2 &    2 &   30 &    6 &  166 &   75 &  214 &  560 &  985 &  334 &  139 &   44 &    0\\
no overlap$^{c}$ &   63 &   22 &    2 &    0 &    2 &   22 &    0 &   99 &   53 &  150 &  377 &  675 &  223 &   94 &   30 &    0\\
\tableline 
\end{tabular}\\
\end{center}
\scriptsize
\noindent
$^{a}$ internal error also known as instrument sensitivity\\
$^{b}$ binning is from July to June with January at the center of bin.\\ 
$^{c}$ values have been corrected for overlapping target lists and for estimated numbers 
of single line spectroscopic binaries (SB1). \\

\end{table}


 
\section{Appendix A: Estimating Target List Overlap}
\label{app:total}
To compute the total number of target stars being monitored (Fig.~\ref{fig:EP_Hist_pyramid})
we need to avoid double-counting.
Target lists of six of the eight surveys considered here have been published, 
thus allowing the overlap in the FGK targets of these six to be eliminated by comparing target star names
(Table 3). 
The total number monitored by these six surveys is then known ($N_{known} \approx 1124$). 
The total number of targets monitored by the Coralie and Elodie surveys is 
known ($N_{C}= 1100, N_{E}= 350$)   
but without published target lists the extent of overlap with the other 
surveys can only be estimated.  We do this by using the statistics of duplicate detections. Let the total number
of exoplanet hosts discovered or confirmed by Elodie and Coralie be $N_{Ehosts}= 16$ and $N_{Chosts}= 33$ respectively.
The number of these planet hosts that were discovered or confirmed by any of the other 6 surveys are
$N_{Eoverlap}= 8$ and $N_{Coverlap}=10$. Since there is no overlap between the Coralie and Elodie surveys 
(Udry private communication), these two sets of planet hosts are mutually exclusive.
The fractional overlap of detections/confirmations is $g_{E} = N_{Eoverlap}/N_{Ehosts} = 8/16 = 0.50$ and 
$g_{C} = N_{Coverlap}/N_{Chosts}= 10/33 = 0.30$  (Tables 3 \& 4). 
We use these fractional detection overlaps as estimates of the fractional target list overlaps.
Thus, we estimate the total number of monitored targets (excluding overlap) as:
\begin{eqnarray}
N_{total} \approx  N_{known} + (1 - g_{C}) \; N_{C}  +(1 - g_{E}) \; N_{E}. 
\end{eqnarray}
However, estimated this way, $N_{total}$ includes target stars that were found to be single line
spectroscopic binaries (SB1).  Exoplanet detection is difficult in such systems so we correct for this by 
substracting the estimated fractions of SB1's in the various surveys (between $3\%$ and $9\%$, see Tables 3 \& 4).  
After this last step we find that $N_{total} \approx 1812 \pm 103$.
The time dependence of $N_{total}$ (Fig. \ref{fig:EP_Hist_pyramid}, top panel)
is taken from the time dependence of $N_{known}$ and from $N_{C}$ and $N_{E}$
(Table 3)       
using the approximation that $g_{C}$ and $g_{E}$ are constants.
Our estimates of the error associated with this procedure are indicated by the error bars on each bin in 
Figs. \ref{fig:EP_Hist_pyramid} \& \ref{fig:EP_Hist_timeline}. 


\section{Appendix B: $\msin  - P$ to $K-P$ parameter space}

We can use the fraction $\approx 11\%$ 
at a duration of $15$ years in the lower panel of Fig. \ref{fig:EP_Hist_timeline} as well as the
known quantities  $P_{max}$, $K_{min}$ to perform 
an interesting consistency check of the relationships 
between them and the extracted values of $\alpha$ and $\beta$. 
Assuming $M_{host} \approx M_{\odot} >> M$ and low eccentricity, the relation between  $\msin$ and the semiamplitude $K$ of
radial velocity is 
\begin{eqnarray}
K \approx A \frac{\msin}{P^{1/3}},
\label{eq:KM}
\end{eqnarray}
where $A = 200$ when $K$ is expressed in units of m/s, $\msin$ in $M_{J}$ and $P$ in days. 
We are only interested in exoplanets with $\msin < 13 M_{Jupiter}$ ($\sim$ the deuterium burning limit).
Using Eq. \ref{eq:KM} to change variables from $\msin$ to $K$ we can write,
\begin{equation}
df = c (\msin)^{\alpha} P^{\beta} \; d(\msin) \; dP 
\end{equation}
(Eq. \ref{eq:TT}) as,
%
\begin{eqnarray}
df = \frac{c}{A^{(1 + \alpha)}} K^{\alpha} P^{\beta + \frac{1}{3}(1 + \alpha)}\; dK \;dP.
\end{eqnarray}
Under the assumption that $\msin$ is uncorrelated with $P$ 
and that $\alpha$ and $\beta$ are approximately constant within the region of interest, 
we can integrate between
arbitrary limits. The upper limit on $K$ depends on period: $K_{max}(P) = 13\;A/P^{1/3}$
from Eq. \ref{eq:KM}. We then have,   
\begin{eqnarray}
f = \frac{c}{(1 + \alpha)A^{(1 + \alpha)} } 
\left[  
(13\;A)^{(1 + \alpha)} \int^{P_{max}}_{0} P^{\beta} dP  - 
K_{min}^{(1 + \alpha)} \int^{P_{max}}_{0} 
P^{\beta + \frac{1}{3}(1 + \alpha)} dP
\right]
\label{eq:kmin}
\end{eqnarray}
Using our best estimates  $\alpha = -1.8$, 
$\beta = -0.3$, 
and $c = 8.6 \times 10^{-5}$ and $f= 0.106$  
from the lower panel in Fig. \ref{fig:EP_Hist_timeline} at $15$ years, 
we integrate Eq. \ref{eq:kmin} and solve for $K_{min}$. We find $K_{min} = 16^{+6}_{-4}$ m/s.
The weighted average internal error of the original Lick and McDonald observing programs that contain 
the 85 stars in these two bins is $\sigma = 6 - 10$ m/s. 
The minimum signal to noise of exoplanet detections is $\sim 3$ so we expect
$K_{min} \approx 3\;\sigma$ which is indeed the case.


\end{document}